\documentstyle[12pt]{article}
\begin{document}

\title{On the problem of an electron scattering in an arbitrary
one-dimensional
potential field}
\author{D.M. Sedrakian$^{*}$, A.Zh.Khachatrian$^{**}$ \\
$^{*}${\small Department of Physics, Yerevan State University,}\\
{\small Manoogian 1,Yerevan 375049, Armenia}\\
$^{**}${\small Department of General Physics, Armenian State Engineering
}\\
{\small \ University, Terian 3, Yerevan 375046, Armenia}}
\date{\today}
\maketitle

\begin{abstract}
Recurrent representations for an electron transmission and reflection
amplitudes for a one-dimensional chain are obtained. The linear
differential
equations for scattering amplitudes of an arbitrary potential are found.
\end{abstract}

In this work the electron scattering problem is considered, when the
potential energy has the form \cite{1Lifsh,2hirota,3Thouls,4And,5Prig}.

\begin{equation}
V(x)=\sum\limits_{n=1}^NV_n(x-x_n)  \label{1}
\end{equation}

Here $V(x-x_n)$ is an individual potential, which has located near the
$x_{n}$point. It is suggested, that individual potentials $V(x-x_n)$ do
not have general points.

There have been many studies to investigate electronic properties of
one-dimensional quaziperiodical and disordered systems. It is well known,
that the problem of determination of transmission amplitude (TA) and
reflection amplitude (RA) in the  field of (\ref{1}) type is reduced to
the
problem of $N$ matrixes product calculation \cite{6abr,7erd,8azb};

\begin{equation}
\left( 
\begin{array}{c}
T_N \\ 
0
\end{array}
\right) =\prod_N^{n=1}\left( 
\begin{array}{cc}
1/t_n^{*} & -r_n^{*}/t_n^{*} \\ 
-r_n/t_n & 1/t_n
\end{array}
\right) \left( 
\begin{array}{c}
1 \\ 
R_N
\end{array}
\right)  \label{2}
\end{equation}
where $T_N$, $R_{N}$are TA and RA of potential $V(x)$, $t_n$ and
$r_{n}$are TA and RA of the individual potential $V(x-x_n)$ of the
system (\ref{1}).

The problem of calculation of the $N$ two-order matrixes product
(\ref{2}),
in general, is equivalent to the problem of solution of some linear
equations \cite{9dan}. So, the problem is represented as

\begin{equation}
D_N=r_N/t_NP_{N-1}+1/t_ND_{N-1}N\geq 1
\label{3.a}
\end{equation}
\begin{equation}
P_N=r_N^{*}/t_N^{*}D_{N-1}+1/t_N^{*}P_{N-1}.N\geq 1  \label{3.b}
\end{equation}
Here $D_N=1/T_N$, $P_N=R_N^{*}/T_N^{*}$ and $D_{N-n}=1/T_{N-n}$, $%
P_{N-n}=R_{N-n}^{*}/T_{N-n}^{*}$, where $T_{N-n}$ and $R_{N-n}$are TA and
RA of the first potentials of the field (\ref{1}). Note, that
in
(\ref{3.a}, \ref{3.b}) $N$ is a variable.

Let us now introduce (\ref{3.a}, \ref{3.b}) in more confortable form.
Excluding $P_{N-1}$ from the first equation and $D_{N-1}$ from the second
one, we will get two equations for $D_N$ and $P_N$ ;

\begin{equation}
D_N=A_ND_{N-1}-B_ND_{N-2}N\geq 2  \label{4.a}
\end{equation}
\begin{equation}
P_N=A_N^{*}P_{N-1}-B_N^{*}P_{N-2},\,\,\,\,\,\,N\geq 2  \label{4.b}
\end{equation}
where $A_N=1/t_N+B_N/t_{N-1}^{*}$ and $B_N=r_Nt_{N-1}/r_{N-1}t_N$.

So the problem of determination of $T_{N}$and $R_{N}$ is
reduced to the problem of solution of the equations (\ref{4.a},
\ref{4.b}).
In order to have the simple solution of (\ref{4.a}, \ref{4.b}), it is
necessary to give initial conditions for equations (\ref{4.a}) and
(\ref{4.b}%
). They are

\begin{equation}
D_1=t_1^{-1},D_0=1,P_1=r_1^{*}/t_1^{*},P_0=0.  \label{5}
\end{equation}

Note, that representations (\ref{4.a}, \ref{4.b}) take place for arbitrary
potentials of the type (\ref{1}). So, if we take $t_N=1+i\alpha _N/2k$ and
$%
r_N=-i(\alpha _N/2k)\exp 2ikx_N$, then (\ref{4.a}) will coincide with the
result of \cite{10gas}, received by the ''determinant'' method for $\delta
$%
- potentials. In a case of layered system from homogeneous mediums, (\ref
{4.a}, \ref{4.b}) yields to the result of \cite{11aron}.

It is interesting to apply the obtained result (\ref{4.a}, \ref{4.b}) to a
simple case, when potentials $V(x-x_n)$ are located periodically and have
the same form. Then $t_{n}$and $r_n$ $\left( n=1,2,...,N\right) $
can be represented as $t_{n}=t_1$ and $r_n=r_1\exp 2ik(x_n-x_1)$ .
In this case, according to (\ref{4.a}, \ref{4.b}), it is clear that
coefficients $A_N$ and B$_{N}$become independent from $N$ and the
equations (\ref{4.a}, \ref{4.b}) are easily solved \cite{12Sed}.

Taking into account conditions (\ref{5}) and denoting the systems period
by $%
a$, for $D_N$ and $P_N$ the following expressions are obtained

\begin{equation}
D_N=\exp iNka\;\left\{ \cos N\beta +iIm(t_1^{-1}\exp (-ika))\sin N\beta
/\sin \beta \right\} ,  \label{6.a}
\end{equation}

\[
P_N=r_1^{*}/t_1^{*}\exp -i(N-1)ka\;\sin N\beta /\sin \beta \;. 
\]
Here $k^2=E\;(\hbar =2m_0=1)$ is the electron energy. The electron energy
spectrum is given by $\cos \beta =Re(t_1^{-1}\exp (-ika))$ , which was
obtained in \cite{13cos}.

Let us now considered one important form of the potential $V(x)$
(\ref{1}),
when $V(x-x_n)$ the are rectangular potentials with arbitrary width $2d_n$
and magnitudes of potentials $V_n$. Then $t_{n}$and $r_n$ are
given
by well-known formulas \cite{14lan};

\begin{equation}
t_n^{-1}=\exp i2kd_n\;\left\{ \cos 2kd_n-i\frac{k_n^2+k^2}{2k_nk}\sin
2kd_n\right\} ,  \label{7.a}
\end{equation}
\begin{equation}
r_n/t_n=i\exp i2kx_n\;\frac{k_n^2-k^2}{2k_nk}\sin 2kd_n.  \label{7.b}
\end{equation}
where $k_n=\sqrt{E-V_n}$, $x_n$ is a coordinate of the middle point of the
potential.

This model (\ref{7.a}, \ref{7.b}) can be used to obtain differential
equations for $D(x)=1/T(x)$ and $P(x)=R^{*}(x)/T^{*}(x)$ quantities, where
$%
T(x)$ and $R(x)$ are TA and RA of the potential

\begin{equation}
V(x)=U(y)\theta (x-y)  \label{8}
\end{equation}
where $U(y)$ is an arbitrary function limited within some interval $a\leq
y\leq b$ and is equal to zero outside of that.

Indeed, approximating $V(x)$ by rectangular potentials, it is possible
with
the help of representation (\ref{4.a}, \ref{4.b}) and corresponding limit
transition to get the sought equation. If we denote in (\ref{4.a},
\ref{4.b}%
) $D_{N-1}=D(x)$, then $D_N=D(x+\Delta x)$ will correspond to the
potential $%
V(x+\Delta x)$. Respectively, $D_{N-2}=D(x-\Delta x)$ will correspond to
the
potential $V(x-\Delta x)$. Presenting in the expression (\ref{7.a},
\ref{7.b}%
,) $V_N=V(x+\Delta x),V_{N-1}=V(x),2d_n=\Delta x$ and expanding
(\ref{4.a}, 
\ref{4.b}) in a series by infinitely small quantity $\Delta x$, we get

\begin{equation}
\frac{1+B}2\frac{d^2D}{dx^2}(\Delta x)^2+(1-B)\frac{dD}{dx}\Delta
x-(A-B-1)D+\theta \left( (\Delta x)^3\right) =0,  \label{9}
\end{equation}
where

\[
B-1=\left( 2ik+\frac 1{V(x)}\frac{dV(x)}{dx}\right) *\Delta x+\theta
\left(
(\Delta x)^2\right) , 
\]
\begin{equation}
A-B-1=V(x)(\Delta x)^2+\theta \left( (\Delta x)^3\right) .  \label{10}
\end{equation}
Inserting (\ref{10}) in to (\ref{9}) and taking $\Delta x\rightarrow 0$,
we
find the desired equation for $D(x)$:

\begin{equation}
\frac{d^2D}{dx^2}-\left( 2ik+\frac 1{V(x)}\frac{dV(x)}{dx}\right)
\frac{dD}{%
dx}-V(x)D=0  \label{11}
\end{equation}
To find TA and RA of the potential $V(x)$ it is necessary to solve the
equation (\ref{11}) with the following initial conditions:

\begin{equation}
D(a)=1,dD/dx\mid _{x=a}=iV(a)/2k  \label{12}
\end{equation}
The value of the function $D(x)$ in the point $x$ determines TA of the
potential $V(x)$. According to the equations (\ref{4.a}, \ref{4.b}), the
quantity $P^{*}(x)$ satisfies the obtained equation (\ref{11}) too. The
initial conditions for $P^{*}(x)$ are the following

\begin{equation}
P^{*}(a)=0,dP^{*}/dx\mid _{x=a}=iV(a)/2k\exp i2ka  \label{13}
\end{equation}

Let us now consider the obtained result (\ref{11}) for a case of a simple
rectangular potential wall with width $L=b-a$ and a constant potential $%
V(x)=V_n$. Then, it is easy to see, that the solution of the problem (\ref
{11})-(\ref{12}) is

\begin{equation}
D(b)=\exp ikL\;\left\{ \cos k_nL-i\frac{k_n^2+k^2}{2k_nk}\sin k_nL\right\}
,
\label{14}
\end{equation}
which, as was expected, agrees with (\ref{7.a}) for $L=2d_n$.

In conclusion, it is interesting to note, that taking $\exp -ikxD(x)$ as
$%
F_1(x)$ and $\exp -ikxP(x)$ as $F_2(x)$, from equation (\ref{11}), we
obtain
the Schr\"{o}dinger equation for function $F_1(x)-F_2(x)=L(x)$;

\begin{equation}
\frac{d^2L}{dx^2}+\left( E-V(x)\right) L=0  \label{15}
\end{equation}
while the $F_1(x)+F_2(x)=\frac ik\frac{dL(x)}{dx}$.

Thus, the problem of determination of TA and RA is reduced to the Cauchy
problem for Shrodinger equation (\ref{15}), with initial condition
(\ref{5}).

\end{document}